\documentclass{article}

\usepackage{arxiv}

\usepackage[utf8]{inputenc} 
\usepackage[russian]{babel}
\usepackage[T1]{fontenc}    
\usepackage{hyperref}       
\usepackage{url}            
\usepackage{booktabs}       
\usepackage{amsfonts}       
\usepackage{nicefrac}       
\usepackage{microtype}      
\usepackage{lipsum}
\usepackage{microtype}      
\usepackage{lipsum}
\usepackage{amsmath}
\usepackage{graphicx}
\usepackage{epstopdf}
\usepackage{cite}

\title{Фоточувствительность и отражательная способность активного слоя в органическом солнечном элементе основанном на таммовском плазмон-поляритоне}

\author{Р.~Г.~Бикбаев\\
Институт физики им. Л.В. Киренского СО РАН, Красноярск, Россия \\
Сибирский федеральный университет, Красноярск, Россия \\
\And
С.~Я.~Ветров\\
Институт физики им. Л.В. Киренского СО РАН, Красноярск, Россия \\
Сибирский федеральный университет, Красноярск, Россия \\
\And
И.~В.~Тимофеев\\
Институт физики им. Л.В. Киренского СО РАН, Красноярск, Россия \\
Сибирский федеральный университет, Красноярск, Россия \\
\And
В.~Ф.~Шабанов\\
Институт физики им. Л.В. Киренского СО РАН, Красноярск, Россия \\
Сибирский государственный университет науки и технологий \\ имени академика М.Ф. Решетнева, Красноярск, Россия \\
}

\begin{document}
\maketitle

\begin{abstract}
Предложена модель органического солнечного элемента, в которой фоточувствительный слой принимает участие в формировании таммовского плазмон-поляритона, локализованного на её границе с многослойным зеркалом.
Показано, что при конструировании таких солнечных элементах можно полностью отказаться от использования металлических контактов, что позволяет избежать нежелательных потерь в системе.
Установлено, что интегральное поглощение в активном слое может быть увеличено на 10\% по сравнению с оптимизированным планарным солнечным элементом.
\end{abstract}


\section{Введение}
Органические солнечные элементы (ОСЭ) на основе сопряженных полимеров привлекают к себе все большее внимания в связи с их низкой стоимостью, легкостью изготовления, малым весом и механической гибкостью солнечных панелей, полученных с помощью технологий рулонной печати~\cite{Atwater2010PlasmonicsDevices,Heeger201425thOperation,He2015Single-junctionPhotovoltage}. 
Подобные ОСЭ содержат объемный гетеропереход, вследствие чего идет поиск компромисса между эффективностями поглощения фотонов и транспортировки носителей заряда. 
Толщина фоточувствительного слоя (ФЧС), в этом случае, составляет не больше 100 нм, что значительно ограничивает эффективность поглощения падающего света. 
В связи с этим, широкое распространение получили методы манипуляции светом для увеличения поглощения в фоточувствительном слое. 
Введение периодических или случайных структур в ФЧС или в интерфейсы ОСЭ приводит к перераспределению оптического поля в них и усилению поглощения фотонов за счет внутреннего рассеяния или эффекта плазмонного резонанса.
Так в 2009 году была экспериментально продемонстрирована возможность увеличения поглощения света в фотоактивном слое, содержащем наночастицы серебра~\cite{Duche2009ImprovingContribution}.
В следующем году в оптически тонких пленках poly-3-hexylthio-phene-phenyl-C61-butyric (P3HT:PC61BM) с плазмонно-резонансными серебряными нанопризмами наблюдалось трехкратное усиление генерации носителей заряда~\cite{P.Kulkarni2010Plasmon-EnhancedNanoprisms}.
Для формирования более широкой линии поглощения в фотоактивном слое было предложено использование наночастиц  различной формы~\cite{Li2013EfficiencyNanoparticles}.
В последствие это направление получило широкое развитие, благодаря бурному развитию технологий изготовления наномасштабных объектов.
Было продемонстрировано, что эффективность ОСЭ может быть повышена за счет внедрения в ФЧС нанопроводов~\cite{Kim2011SilverApplications}, наностержней~\cite{He2015Single-junctionPhotovoltage}, частиц в форме кубов, додекаэдров, октаэдров и треугольных пластин~\cite{Tseng2015Shape-DependentEffect}.
Не менее важным способом повышения эффективности ОСЭ является структурирование границы раздела металл-полупроводник, приводящее к возбуждению поверхностного плазмон-поляритона.
В таких структурах металлические контакты принимают участие не только в сборе носителей заряда, но и при увеличении их концентрации. 

Другим перспективным направлением является внедрение в ОСЭ 1D фотонных кристаллов (ФК). 
Расположение ФК за металлическим контактом приводит к тому, что практически 100\% излучения, падающего на него, отражается и проходит через активный слой вторично, тем самым увеличивая эффективность ОСЭ~\cite{Yu2012SemitransparentCrystals,Yu2013SimultaneousCrystals,Yu2014LightReflectors,Lunt2011TransparentApplications}.
В этом случае открывается возможность использования более тонких металлических пленок в качестве контактов и, как следствие, уменьшения потерь в ОСЭ.
Замена прозрачного контакта ФК-структурой приводит к тому, что активных слой остается между металлической пленкой и ФК.
Как известно, в подобных структурах могут возбуждаться таммовские плазмон-поляритоны~\cite{Kaliteevski2007}, на длине волны которых возникают дополнительные линии поглощения излучения в активном слое.
Этот механизм увеличения эффективности ОСЭ был продемонстрирован в работе~\cite{Zhang2012}.
Автор показал, что благодаря возбуждению таммовского плазмон-поляритона (ТПП), интегральное поглощение в активном слое может быть увеличено с 26.3\% до на 35.3\% в исследуемом интервале длин волн.
В рассмотренных моделях фоточувствительный слой является пассивным поглощающим элементом, не принимающим участие в формировании локализованных состояний.

Новой является идея использования допированного фоточувствительного слоя в качестве зеркала, ограничивающего одномерный фотонный кристалл. 
В этом случае, на их границе раздела локализуется ТПП~\cite{Vetrov2013,Vetrov2017,Bikbaev2019Epsilon-Near-ZeroPolariton}, что приводит к возникновению дополнительной полосы поглощения падающего на структуру излучения и, как следствие, увеличению эффективности ОСЭ. Привлекательность такой структуры заключается в том, что можно полностью отказаться от металлического контакта, обеспечив поглощение только в ФЧС слое.  

\section{Описание модели}

\begin{figure}
\centering
\includegraphics[scale=0.45]{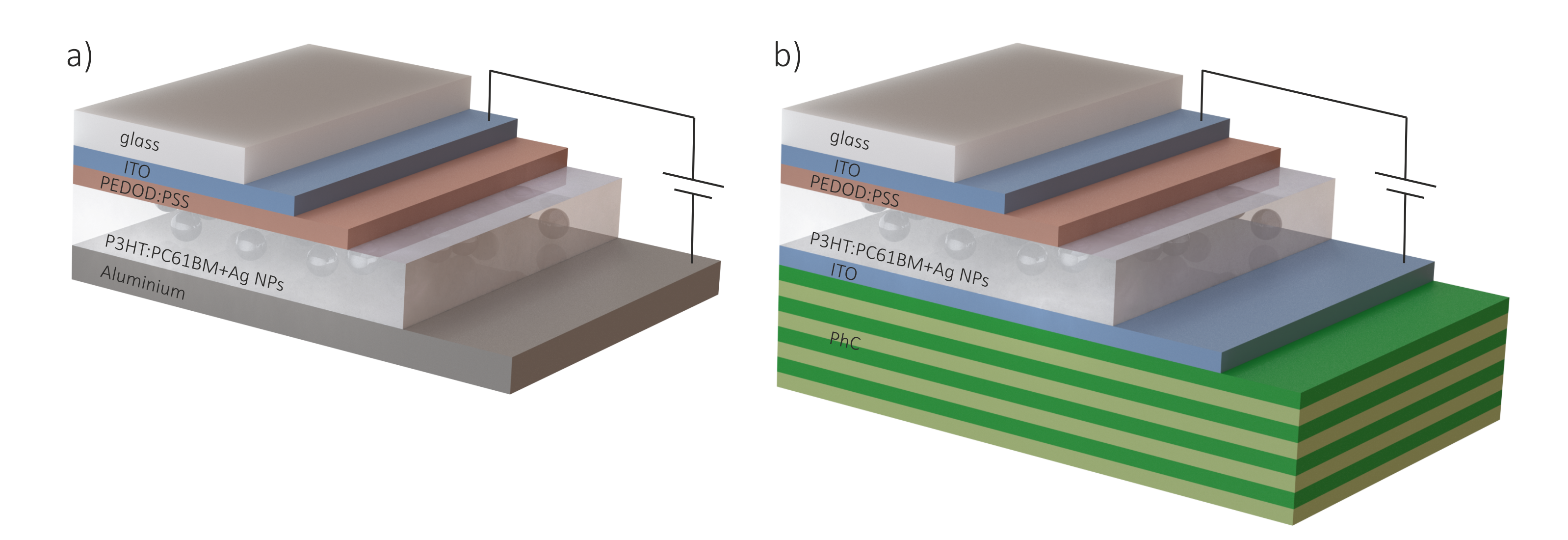}\\
\caption{ (a) Схематичное изображение органического солнечного элемента, допированного плазмонными наночастицами и  (b) схематическое изображение солнечного элемента на основе таммовского плазмон-поляритона. }
\label{fig:fig1}
\end{figure}

Модель органического солнечного элемента, предложенного нами, изображена на рисунке~\ref{fig:fig1}b. 
В отличии от ранее изученного солнечного элемента~(Рис.~\ref{fig:fig1}a) \cite{Kim2005RolesDevices,Wu2013UncoveringCells}, в нашей модели фоточувствительный слой ограничивает не металлическую пленку, а одномерный фотонный кристалл.
Увеличение поглощения в фоточувствительном слое, в этом случае, достигается как за счет 100\% отражения излучения от фотонного кристалла, так и при помощи формирования дополнительной полосы поглощения на длине волны ТПП, локализованного на границе ФЧС-ФК.
Слой P3HT:PC61BM+AgNPs~\cite{Stelling2017PlasmonicCells} толщиной 50~нм, допирован серебряными наношарами с объемной концентрацией 20$\%$. Толщина слоя poly(3,4-ethylenedioxythiophene) polystyrene sulfonate  (PEDOT:PSS)~\cite{Chen2015OpticalCells} равна 20~нм. 
В качестве контактов использованы пленки ITO с толщинами 15~нм и 45~нм. 
Элементарная ячейка фотонного кристалла сформирована из двуокиси кремния (SiO$_2$) и диоксида титана (TiO$_2$) с толщинами $d_{SiO_2}=75$~нм, $d_{TiO_2}=40$~нм и диэлектрическими проницаемостями $\varepsilon_{SiO_2}=1.45$ и $\varepsilon_{TiO_2}=2.4$, соответственно.

Зависимости действительной и мнимой частей показателя преломления P3HT:PC61BM от длины волны падающего излучения изображены на рисунке \ref{fig:fig2}a.
Из рисунка видно, что мнимая часть показателя преломления принимает максимальное значение в интервале длин волн от 350 до 600~нм и, как следствие, падающее на структуру излучение будет поглощаться ею именно в этом спектральном диапазоне. 
В длинноволновой области спектра поглощение близко к нулю.

Эффективная диэлектрическая проницаемость фоточувствительного слоя P3HT:PC61BM+AgNPs, допированного плазмонными наночастицами, определяется с помощью модели эффективной среды~\cite{Maxwell-Garnett1906237}:
\begin{equation}
\varepsilon_{\text{eff}}=\varepsilon_d\left[1+\frac{f\left(\varepsilon_m(\omega )-\varepsilon
_{d}\right)}{\varepsilon_{d}+\left(1-f\right)(\varepsilon_m(\omega )-\varepsilon_{d})1/3}\right],
\label{eq:Eq1}
\end{equation}
где $f$~–~фактор заполнения, то есть, объемная доля наночастиц в матрице;
$\varepsilon_{d}$ и $\varepsilon_m$($\omega $)~–~диэлектрические проницаемости соответственно матрицы и металла, из
которого изготовлены наночастицы; $\omega $~–~частота излучения.
Диэлектрическую проницаемость металла, из которого изготовлены наночастицы, найдем, используя приближение Друде: 
\begin{equation}
\varepsilon_{m}(\omega)=\varepsilon _{{\infty}}-\frac{\omega _p^2}{\omega ^2+i\omega \gamma },
\label{eq:Eq2}
\end{equation}
где $\varepsilon_0$ – постоянная, учитывающая вклады межзонных переходов связанных электронов;
$\omega_p$ – плазменная частота;
$\gamma$ – величина, обратная времени релаксации электронов.
Для серебра $\varepsilon_0$ = 5, $\omega_p$ = 9 эВ и $\gamma$ = 0.02 эВ.

\section{Результаты расчета}

\begin{figure}
\centering
\includegraphics[scale=0.7]{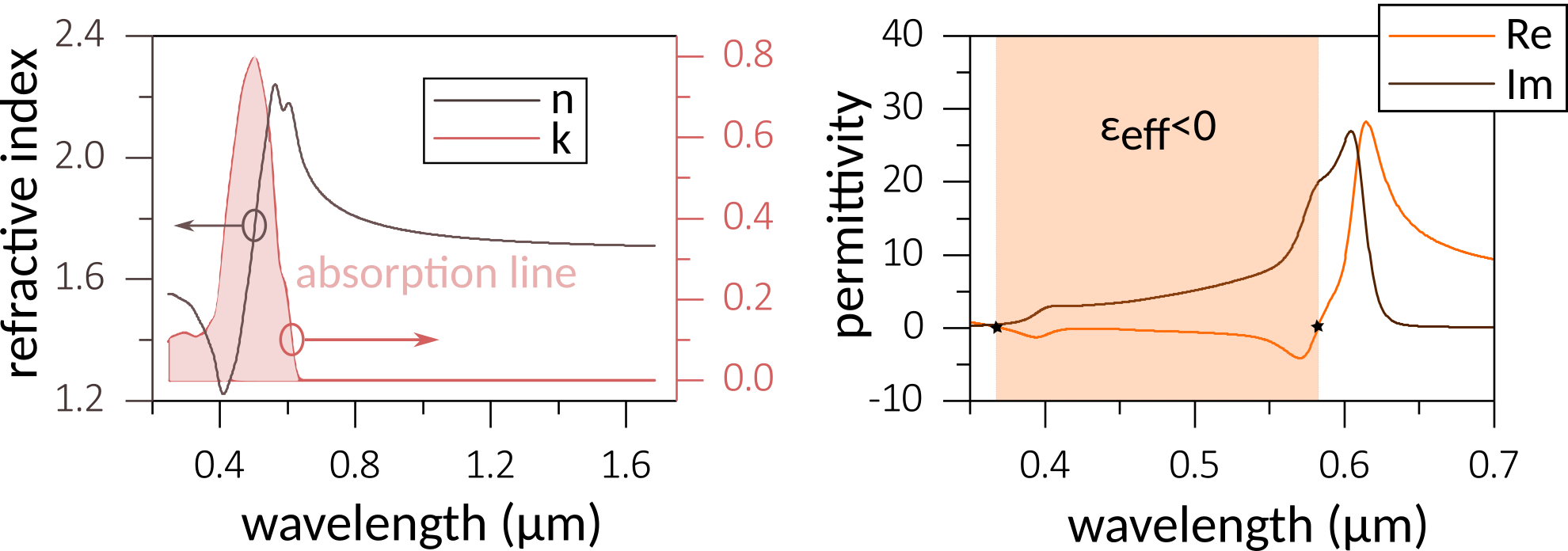}\
\caption{(а) Зависимости действительной и мнимой части показателя преломления P3HT:PC61BM  и (b) зависимости действительной и мнимой части эффективной диэлектрической проницаемости  фоточувствительного слоя, допированного плазмонными наночастицами,  от длины волны падающего излучения.}
\label{fig:fig2}
\end{figure}

\begin{figure}
\centering
\includegraphics[scale=0.7]{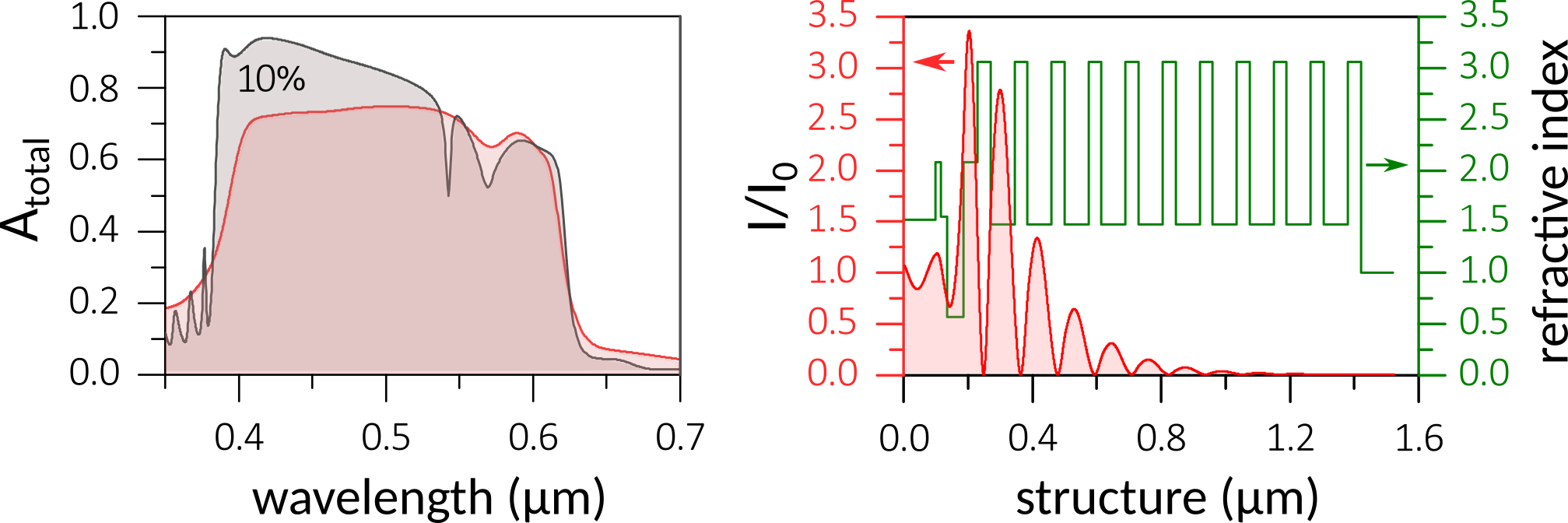}\
\caption{(а) Интегральное поглощение в фоточувствительном слое в структурах ITO/PEDOT:PSS/P3HT:PC61BM+AgNPs/Al и ITO/PEDOT:PSS/P3HT:PC61BM+AgNPs/ITO/PhC и (b) пространственное распределение показателя преломления в структуре и локальной интенсивности поля на длине волны ТПП.}
\label{fig:fig22}
\end{figure}

Зависимости действительной и мнимой части эффективной диэлектрической проницаемости, рассчитанные по формуле~\ref{eq:Eq1}, от длины волны падающего излучения изображены на рисунке~\ref{fig:fig2}б.
Из рисунка видно, что мнимая часть ДП достигает максимального значения на длине волны 600 нм, что обусловлено плазмонным резонансом в наночастицах. 
Также отметим, что действительная часть принимает отрицательные значения в широком интервале длин волн (от 360 до 570~нм) и, как следствие, в этом интервале активный слой выступает в качестве металлического зеркала.
Сопряжение активного слоя, обладающего подобными спектральными характеристиками, с фотонным кристаллом приведет к формированию на их границе раздела таммовского плазмон-поляритона.  
Для подтверждения этого факта методом трансфер-матрицы~\cite{Yeh1979} был произведен расчет интегрального поглощения в слое P3HT:PC61BM+AgNPs для классического планарного солнечного элемента и солнечного элемента, сопряженного с ФК.
Результаты расчета изображены на рисунке~\ref{fig:fig22}а.
Под интегральным поглощением мы подразумеваем поглощение в фоточувствительном слое, нормированный на спектр солнечного излучения.
В общем случае, интегральное поглощение определяется отдельно для каждой поляризации $A_{TE}$ и $A_{TM}$, а их среднее арифметическое $A_{total}=(A_{TE}+A_{TM})/2$ дает полное поглощение в слое. 
В случае нормального падения $A_{total}= A_{TE} = A_{TM}$  и может быть определено как:

\begin{equation}
A_{total} = \frac{\int\limits^{\lambda_2}_{\lambda_1} A(\lambda)S(\lambda)\,d\lambda}{\int\limits^{\lambda_2}_{\lambda_1} S(\lambda)\,d\lambda},  
\end{equation}

где $\lambda_1=350$~nm, $\lambda_2=700$~nm, $A(\lambda)$ -- поглощение в слое P3HT:PC61BM+AgNPs, $S(\lambda)$ -- спектр солнечного излучения (AM1.5) .

Расчеты показали, что в предложенной модели ОСЭ интегральное поглощения в исследуемом интервале длин волн увеличивается на $\approx$10\% (c 50.52\% до 55.36\%), в сравнении с аналогичным планарным ОСЭ~(Рис.~\ref{fig:fig1}a). 
Это достигается за счет формирования таммовского плазмон-поляритона, локализованного на границе фотонного кристалла и активного слоя, допированного плазмонными наночастицами. 
На длине волны ТПП ($\lambda_{TPP}=400$~нм), эффективная диэлектрическая проницаемость фоточувствительного слоя равна $\varepsilon_{\text{eff}} = -1.007+2.716i$. Пространственное распределение поля на длине волны ТПП изображено на рисунке~\ref{fig:fig22}b. 

Из рисунка видно, что поле локализовано на границе ФК и фоточувствительного слоя и экспоненциально затухает по обе стороны от их границы раздела. 
При этом поле локализовано в области соизмеримой с длиной волны света.
Важно отметить, что интенсивность поля на длине волны ТПП лишь в 3.5 раза больше интенсивности падающего поля.
Столь незначительное усиление может быть объяснено формированием на границе ФЧС-ФК таммовского плазмон-поляритона с широкой спектральной линией.
Формирование такого рода состояний возможно в системах с большими потерями, как это было продемонстрировано в работе~\cite{Vyunishev2019BroadbandPolariton}.
В нашей структуру мы имеем аналогичную ситуацию, так как действительная часть эффективной диэлектрической проницаемости ФЧС на длине волны ТПП практически в 3 раза меньше её мнимой части.

\section{Заключение}
В работе предложена модель органического солнечного элемента, в котором фоточувствительный слой выступает не только в роли поглотителя, но и зеркала, принимающего участие в формировании локализованного состояния. 
Энергетические спектры структуры и распределение локальной интенсивности в ней рассчитаны методом трансфер матрицы.
Показано, что в предложенной модели интегральное поглощение в фоточувствительном слое увеличивается на 10\% в сравнении с ранее предложенной моделью ОСЭ.  
Установлено, что увеличение поглощения в этом случае достигается за счет формирования дополнительной полосы поглощения в ОСЭ, и обусловлено формированием на границе фоточувствительного слоя и фотонного кристалла таммовского плазмон-поляритона.

\bibliographystyle{unsrt}
\bibliography{references}

\end{document}